\documentclass[
 reprint,amsmath,amssymb,aps,floatfix,]{revtex4-2}

\usepackage{physics}
\usepackage{graphicx}
\usepackage{dcolumn}
\usepackage{bm}
\usepackage{hyperref}

\usepackage{color}

\usepackage{siunitx}
\DeclareSIUnit\gauss{G}
\DeclareSIUnit\g{\Gamma}
\DeclareSIUnit\uK{\micro\K}
\DeclareSIUnit\nK{\nano\K}
\DeclareSIUnit\Tf{T_{F}} 
\DeclareSIUnit\isat{I_{sat}}
\newcommand{\rb}{\ifmmode ^{87}{\text{Rb }} \else $^{87}{\text{Rb }}$\fi}
\newcommand{\kk}{\ifmmode ^{40}{\text{K }} \else $^{40}{\text{K }}$\fi}
\newcommand{\done}{\ifmmode \mathcal{D}1 \else $\mathcal{D}1$ \fi}
\newcommand{\dtwo}{\ifmmode \mathcal{D}2 \else $\mathcal{D}2$ \fi}

\usepackage{tabularx}
\begin{document}

\preprint{APS/123-QED}
\title{Two-colour laser cooling for \kk-\rb quantum gas mixtures}
\author{Yann Kiefer, Max Hachmann, Andreas Hemmerich}
\email{hemmerich@physnet.uni-hamburg.de}
\affiliation{Institut f{\"u}r Quantenphysik, Universit{\"a}t Hamburg, 22761 Hamburg, Germany \\
Zentrum f{\"u}r optische Quantentechnologien, Universit{\"a}t Hamburg, 22761 Hamburg, Germany \\
The Hamburg Centre for Ultrafast Imaging, 22761 Hamburg, Germany}

\date{\today}

\begin{abstract}
We present an efficient cooling scheme for fermionic \kk atoms, using laser light red and blue detuned with respect to the $\mathcal{D}2$ and $\mathcal{D}1$ principle flourescence lines, respectively. The cooling scheme is found to significantly increase the saturation level for loading of a \kk magneto-optical trap (MOT), resulting in increased atom numbers or decreased cycle times. While the attainable \kk atom number is approximately doubled if exclusively \kk atoms are cooled, the scheme is particularly powerful for dual-species MOTs, for example, if \kk and \rb atoms are cooled simultaneously in the same MOT configuration. The typical atom losses due to light-assisted hetero-nuclear collisions between \kk and \rb seem to be reduced giving rise to a threefold improvement of the \kk atom number as compared to that in a conventional dual-species MOT, operating merely with $\mathcal{D}2$ light. Our scheme can be a useful extension to most dual-species experiments, aiming to reach simultaneous degeneracy of both species.
\end{abstract}

\maketitle

\section{\label{sec:Introduction}Introduction}
The recent advances in quantum simulation platforms based on ultracold neutral atoms \cite{Sherson2010, Mazurenko2017} and molecules \cite{DeMarco2019, Schindewolf2022, Stevenson2023, Kiefer2023} have triggered renewed interest in refined laser cooling techniques enabling rapid cycle times and high fidelity sample preparation\cite{Bernien2017, Brown2023}. To study fundamental low-temperature phases of matter \cite{Schafer2020, Chomaz2023}, degenerate quantum gases are typically employed, requiring temperatures on the 100 nK scale, well below the degeneracy temperature $T_{deg}$, where the behaviour of matter is governed by quantum mechanics. The usual initial step for preparation of a single-species quantum degenerate gas is laser-cooling to temperatures below the Doppler temperature $T_D$ (on the order of $100\,\mu$K) in a three-dimensional (3D) magneto-optical trap (3D-MOT) \cite{Raab1987}, which combines an optical molasses \cite{Dalibard1989} with a 3D magnetic quadrupole field. If more than a single atomic species is required, for example, for the production of ultracold hetero-nuclear molecules, the simultaneous, spatially overlapped operation of multiple 3D-MOTs is necessary referred to as a dual-species 3D-MOT. This typically comes with undesirable density and atom number limitations due to two-body loss processes such as molecule association via light assisted collisions in presence of photons of the cooling light \cite{Anderson1994}. In this article we present a novel cooling scheme in a dual-species 3D-MOT, which mitigates such losses for the specific case of a \kk-\rb mixture \cite{Goldwin2002, Ospelkaus2006}. 

A possible way to decrease such losses is the use of the \textit{Spontaneous Force Optical Trap} (SPOT) technique discussed in Ref.~\cite{Ketterle1993}. This method requires to place a shadow into the repumping beam blocking the light from irradiating the atoms in the trap center. As a result since the cooling cycling transition for alkali atoms is typically not completely closed, the atoms in the trap center naturally accumulate in a dark state and exit the cooling cycle. Hence, the overall fluorescence and the rate of light-assisted collisions decrease. Unfortunately, the SPOT technique is only efficient for rather incompletely closed cycling transitions as in the case of sodium. Hence, it also does not apply for many dual-species mixtures of interest. For single-species scenarios, more recently, laser cooling using multiple atomic transitions was employed and found favourable for improved 3D-MOT performance in experiments of alkaline \cite{Salomon_2013} and alkaline earth isotopes \cite{Hoschele2023}. In these works, the improved performance is linked to shielding effects due to short lived meta-stable intermediate states. 

In the present work, we explore a novel two-colour scheme for cooling a dual-species \kk-\rb mixture, where in addition to the conventional laser radiation, detuned to the red side of the \dtwo-transition of \kk at \SI{767}{\nano\meter}, we include a second spectral component, detuned to the blue side of the \done-transition of \kk at \SI{770}{\nano\meter}. The second species, \rb, is cooled by the conventional single-colour method using red-detuned light on the \dtwo-transition of \rb at \SI{780}{\nano\meter}. After introducing the baseline scenario of a single-colour, single-species \kk 3D-MOT, we benchmark the performance of the two-colour cooling scheme in a single-species \kk 3D-MOT in the saturated regime. Here, we find an approximately twofold improvement of the number of cooled atoms, as compared to the conventional single-colour laser cooling scheme \cite{DeMarco1999}. Subsequently, by adding a conventional single-colour 3D-MOT for \rb, we investigate the dual-species (\kk-\rb) 3D-MOT performance when the novel two-colour cooling scheme is applied, finding an up to threefold improvement of the final potassium atom number, which we attribute to the suppression of light-assisted hetero-nuclear collisions.  
\section{Experimental setup} 
\paragraph{General setup:} The experimental setup consists of a dual-species 3D-MOT formed by the superposition of single-species MOTs for \kk and \rb inside an ultra-high vacuum glass cell. The dual-species 3D-MOT is loaded by cold atomic beams of \kk and \rb atoms obtained from two separated source regions, which connect to the glass cell via differential pumping stages. The preparation of the cold beams is achieved in so-called $\textrm{2D}^+$ - MOTs, loaded from dispensers. Essentially, this is a 2D-MOT using a magnetic field with linear quadrupole geometry and two additional beams along the zero magnetic field axis, which is referred to by the "+" sign in the acronym. A detailed description of this technique is found in Ref.~\cite{Dieckmann1998}. 
\paragraph{$\textrm{2D}^+$ - MOTs:} The operating parameters of the $\textrm{2D}^+$-MOTs can be found in \autoref{tab:table2DK} (\kk) and \autoref{tab:table2DRb} (\rb) and are not varied throughout all experiments presented in this paper. The rubidium $\textrm{2D}^+$- is operated on the $5^2 S_{1/2} \rightarrow 5^2 P_{3/2}$ transition ($\mathcal{D}2$). The frequency of the cooling light is tuned to the red side ($\delta<0$) of the $\ket{F=2}\rightarrow \ket{F'=3}$ transition, while the frequency of the repumping light (required to prepare a closed excitation cycle) is resonant to the $\ket{F=1}\rightarrow \ket{F'=0}$ transition. The potassium $\textrm{2D}^+$-MOT is operated on the $4^2 S_{1/2} \rightarrow 4^2 P_{3/2}$ transition ($\mathcal{D}2$), with the cooling light tuned to the red side of the $\ket{F=9/2}\rightarrow \ket{F'=11/2}$ transition and the repumping light adjusted in resonance with the $\ket{F=7/2}\rightarrow \ket{F'=9/2}$ transition. This is indicated by the red arrows in \autoref{Konly}(a), where the electronic level scheme of \kk is shown. For both species, the circular polarisations of the cooling and repumping light in the $\textrm{2D}^+$-MOTs are orthogonal. The intensity ratios $I_{cool}/I_{rep}$ of the cooling and the repumping light in the $\textrm{2D}^+$-MOTs are $10$ for \rb and $1$ for \kk. 
\paragraph{\rb 3D-MOT:} The \rb 3D-MOT uses the same single-colour cooling scheme as the \rb $\textrm{2D}^+$-MOT. Similarly as the parameters of both $\textrm{2D}^+$ - MOTs, the parameters of the \rb 3D-MOT are not varied for all experiments reported here, in order to reduce complexity.  
\paragraph{\kk 3D-MOT:} To realise the two-colour cooling scheme, the \kk 3D-MOT beams are derived from three tapered amplifiers (TA), divided into six beams of equal intensity delivered by polarisation maintaining optical fibers. The cooling and repumping light on the $\mathcal{D}2$-transition is provided by two separate TAs. The cooling/repumping scheme on the $\mathcal{D}1$-transition is realised by a single TA, where the phase-coherent repumping light is imprinted by an electro-acoustic modulator operating at $f_{HFS,^{40}K}=\SI{1.285}{\giga\hertz}$ \cite{Tiecke2019}. The magnetic quadrupole field centered around the position of the atoms is realised with a pair of water-cooled copper coils placed outside the vacuum system.
%
%
\section{\label{sec:2colourMotPotassium}single-species \kk MOT}

\begin{figure}[t]
\includegraphics[width=\columnwidth]{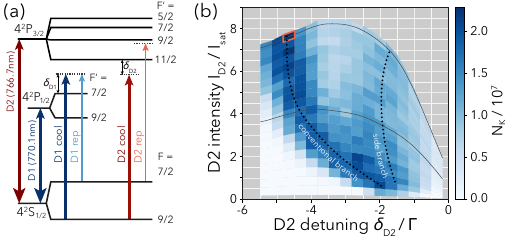}
\caption{In (a), the electronic level scheme of \kk is shown. The thick red (blue) arrows indicate the $\mathcal{D}2$($\mathcal{D}1$) cooling transitions. The thin arrows indicate the respective repumping transitions. The frequencies of the respective transitions are indicated by their detunings from resonance $\delta_{D2}<0$ and $\delta_{D1}>0$. In (b), the \kk atom number, after loading during $t_{load,K}=\SI{3}{\second}$, is plotted against the intensity $I_{D2}$ of the \dtwo cooling light and the \dtwo frequency detuning $\delta_{D2}$. The solid gray lines indicate two intensities $I_{D2}=I_{max}$ (upper) and $I_{D2}=0.5 \,I_{max}$ (lower). Black dotted lines are a guide to the eye to indicate the two regimes, see main text.}
\label{Konly}
\end{figure} 

\begin{figure}[t]
\includegraphics[width=\columnwidth]{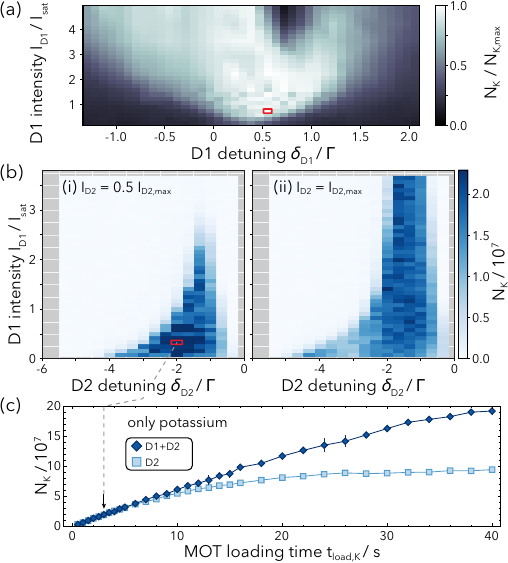}
\caption{In (a), the normalised \kk atom number $N_K/N_{K,max}$ is plotted against the \done cooling light intensity $I_{D1}$ and the cooling light detuning $\delta_{D1}$. In (b), the \kk atom number $N_{K}$ after $t_{load}=\SI{3}{\second}$ is plotted against the \dtwo cooling light detuning $\delta_{D2}$ and the \done intensity $I_{D1}$ for two distinct settings of the \dtwo intensity $I_{D2}=0.5 I_{D2,max}$(i) and $I_{D2}=I_{D2,max}$(ii). In (a) and (b), the red rectangle indicates the parameter settings where maximal \kk atom number is found. In (c), $N_{K}$ is plotted against the loading time $t_{load,K}$. The light blue squares show $N_{K}$ for optimal parameters according to \autoref{Konly}(b), when only the \dtwo-transition is used for cooling. The dark blue diamonds show $N_{K}$ for cooling on both \done- and \dtwo-transitions for optimal parameters denoted by the black arrow and the red square in (b).}
\label{KD1D2}
\end{figure} 

To assess the performance of the two-colour cooling scheme, for comparison, we first characterise the operation of a conventional single-colour, single-species MOT of \kk atoms. In \autoref{Konly}(b), we plot the \kk particle number as a function of the \dtwo cooling light intensity $I_{D2}$ and the \dtwo cooling light frequency detuning $\delta_{D2}$ for a loading time of $t_{load,K}=\SI{3}{\second}$. 

In this plot, we identify two different domains, which are clearly separated for $I_{D2}>\SI{3}{\isat}$.

The first domain (\textit{conventional branch}) is found around $\delta_{D2}<\SI{-4}{\Gamma}$ and $I_{D2}>\SI{3}{\isat}$ (large negative detuning and large intensity) denoted by the black dotted line, which serves as a guide to the eye. Here, we find the maximum particle number of $N_{K}=\num{2.2e7}$ for $\delta_{D2}=\SI{-3.5}{\Gamma}$ and $I_{D2}=I_{D2,max}=\SI{7.5}{\isat}$ (red rectangle in \autoref{Konly}(b)). The second domain (\textit{side branch}) is found for more resonant \dtwo cooling light frequency detuning $\delta_{D2}\sim \SI{-2}{\Gamma}$ at almost all settings of the intensity $I_{D2}$. Here, we find particle numbers up to $N_{K}=$\num{1.5e7}. Between these two domains we observe a local minimum of the particle number $N_{K}$. While for the single-colour scheme, the second domain of the \kk 3D-MOT shows slightly less efficient performance, this domain becomes relevant if the two-colour operation of the MOT is investigated.

In contrast to conventional laser cooling in a MOT, we additionally employ blue-detuned light on the \done-transition of \kk as indicated by the set of blue arrows in \autoref{Konly}(a), similar to the experiments presented in \cite{Salomon2013} and reminiscent of VSCPT \cite{Weidemueller1994} and gray molasses cooling \cite{RioFernandes2012}. In \autoref{KD1D2}(a), we plot the normalised potassium particle number $N_K/N_{K,max}$ as a function of the \done detuning $\delta_{D1}$ and the \done intensity $I_{D1}$. In this experiment we fix the parameters of the \dtwo cooling light to $I_{D2}=\SI{5.76}{\isat}$ and $\delta_{D2}=\SI{-1.67}{\Gamma}$ (optimal values according to \autoref{KRBD1D2}). We identify two regimes separated by a local minimum around $I_{D1}>\SI{3}{\isat}$ and $\SI{0.4}{\Gamma}<\delta_{D1}<\SI{1.0}{\Gamma}$. Our qualitative understanding of these domains of optimal performance is limited and subject of further investigation. The parameter setting for optimal MOT performance of the \done frequency detuning $\delta_{D1}=\SI{0.55}{\Gamma}$ is indicated by the red rectangle in \autoref{KD1D2}(a) and is kept at this value for all following investigations. 

First, we vary the intensity $I_{D1}$ of the \done cooling light and the detuning $\delta_{D2}$ of the \dtwo cooling light for two different settings of the intensity $I_{D2}$ of the \dtwo cooling light, as can be seen in \autoref{KD1D2}(b). The intensity of the \dtwo cooling light is set $I_{D2}=0.5\,I_{D2,max}$ in \autoref{KD1D2}(b)i and to $I_{D2}=I_{D2,max}$ in \autoref{KD1D2}(b)ii according to the solid gray lines in \autoref{Konly}(b). The colour map is set to be identical to the colour map in \autoref{Konly}(b), so that the saturated data points correspond to an improved MOT performance. We observe that with the inclusion of cooling light operating on the \done transition of \kk, the optimal setting of the \dtwo cooling frequency detuning is at $\delta_{D2}=\SI{-2}{\Gamma}$ in contrast to the optimal detuning without cooling on the \done transition, which is found to be $\delta_{D2} = -4.65\, \Gamma$. Additionally, we have varied the ratio of the intensities $I_{D1}/I_{D2}$. If $I_{D2}=I_{D2,max}$ (see \autoref{KD1D2}(b)ii), $I_{D1}$ can be set to larger values and still yield reasonable \kk MOT atom numbers. However, we find the best performance for parameter settings $I_{D2}=0.5\,I_{D2,max}, I_{D1}=\SI{0.33}{\isat}$ and $\delta_{D2}=-2\Gamma$ according to the data point indicated with a red rectangle in \autoref{KD1D2}(b)i. Here, we obtain $N_{K}=\num{2.51e7}$ after $t_{load,K}=\SI{3}{\second}$, a \SI{13}{\percent} improvement compared to the single-colour operation, as indicated by the black arrow in \autoref{KD1D2}(c).

As a last step, we investigate the \kk atom number $N_{K}$ as a function of the MOT loading time $t_{load}$ shown in \autoref{KD1D2}(c). For this, we load the MOT for variable times at optimal parameters indicated by the rectangles both for the \dtwo-only MOT (see \autoref{Konly}(b)) and the two-colour MOT (see \autoref{KD1D2}(b)i). For short loading times of $t_{load,K}<\SI{8}{\second}$ the improvement by the two-colour cooling scheme is marginal. However, for longer loading times $t_{load,K}>\SI{8}{\second}$ an almost twofold increase of the \kk atom number $N_{K}$ is obtained. To extrapolate the saturation levels of the two cooling methods, we fit both data sets in \autoref{KD1D2}(c) with a time dependent rate equation model, including a density dependent two-body loss term \cite{Gruenert2001}, and find a threefold increased saturation level in the two-color scenario. The loading curve is found to be well fitted by this model, indicating that, in contrast to the single-colour MOT, even larger atom numbers can be expected, by increasing the flux  of the cold \kk beam source. See Appendix B for details of the fit model.
\section{\label{sec:2SpeciesMOT}Single-colour dual-species MOT}
\begin{figure}[t]
\includegraphics[width=\columnwidth]{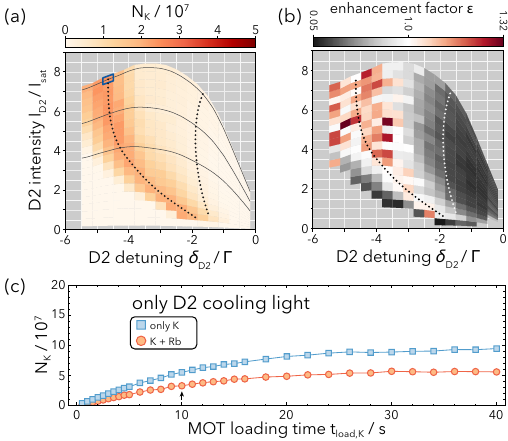}
\caption{In (a), the potassium atom number is plotted as a function of the intensity $I_{D2}$ and the frequency detuning $\delta_{D2}$ in the dual-species MOT in absence of any cooling light operating on the \done transition. In (b), we plot the improvement factor defined as the ratio of the normalised particle numbers $N_K$ found for single - and dual-species MOT operation. In (c), we plot the particle number $N_K$ of the single-species (light blue squares) and the dual-species (orange circles) MOT as a function of the loading time $t_{load,K}$. The blue square in (a), close to the upper left corner, indicates the optimal parameters used for measuring the loading curve in (c). The black arrow in (c) indicates the loading time for which (a) is obtained.}
\label{KRBD2}
\end{figure} 
In this section, we investigate the negative effects of light-assisted hetero-nuclear collisions (LAHNCs) on the performance of a dual-species MOT for \kk and \rb atoms, operating with conventional single-colour \dtwo-only cooling. Since the loading efficiency of rubidium is much larger than that of potassium, we apply different loading times for both species. Typically, we operate with a variable loading time $t_{load,K}$ for \kk atoms followed by a time window with a duration $t_{load,KRb} = \SI{3}{\second}$, during which both species \kk and \rb are loaded. During the combined loading period $t_{load,KRb}$ we see no further increase of the \kk atom number $N_K$ but rather a decrease attributed to LAHNCs. In order to consistently compare loading of \kk atoms with loading of \kk and \rb atoms, we equally specify the \kk atom number $N_{K}$ and the loading time $t_{load,K}$ for both cases.

In \autoref{KRBD2}(a), we plot the \kk atom number $N_{K}$ as a function of the intensity  $I_{D2}$ and the detuning $\delta_{D2}$ of the \dtwo cooling light. In the \textit{conventional branch}, we observe a similar qualitative behaviour as previously found in \autoref{Konly}(b)) for the single-species \dtwo-only MOT. In \autoref{KRBD2}(a), the maximum particle number $N_{K}=\num{2.38e7}$ is found in the \textit{conventional branch} for a detuning $\delta_{D2}=\SI{-4.6}{\Gamma}$ and an intensity of $I_{D2}=\SI{7.7}{\isat}$. However, the \textit{side branch} obtained in \autoref{Konly}(b), vanishes. We explain this behaviour with an increased photon scattering for cooling light tuned closer to the atomic resonance ($\delta_{D2} \sim 0$), which in consequence leads to pronounced LAHNCs and associated two-body loss. 

To further illustrate the effect, we plot the improvement factor $\varepsilon$ in \autoref{KRBD2}(b), which is defined as the ratio of the normalised \kk atom numbers for the single-species (\autoref{Konly}) and the dual-species (\autoref{KRBD2}(a)) single-colour \kk MOT, i.e., both in absence \done cooling. The red shaded areas in \autoref{KRBD2}(b) (improvement factor $\varepsilon > 1$) indicate regions, where the dual-species MOT performs better compared to the single-species case, whereas the black areas ($\varepsilon < 1$) denote the regions where the $N_{K}$ is severely reduced by the presence of \rb atoms. 

In \autoref{KRBD2}(c), we investigate $N_K$ as a function of the loading time $t_{load,K}$. The orange circles denote $N_{K}$ for optimal parameters indicated by the blue rectangle in \autoref{KRBD2}(a). For comparison, the blue squares repeat $N_K$ found in \autoref{KD1D2}(c) for the single-species single-colour scenario. We see, that the \kk atom number for the single-colour dual-species MOT is clearly reduced due to the presence of \rb atoms. The black arrow in \autoref{KRBD2}(c) marks the loading time $t_{load,K} = \SI{10}{\second}$, for which \autoref{KRBD2}(a) is obtained.
\begin{figure}[t]
\includegraphics[width=\columnwidth]{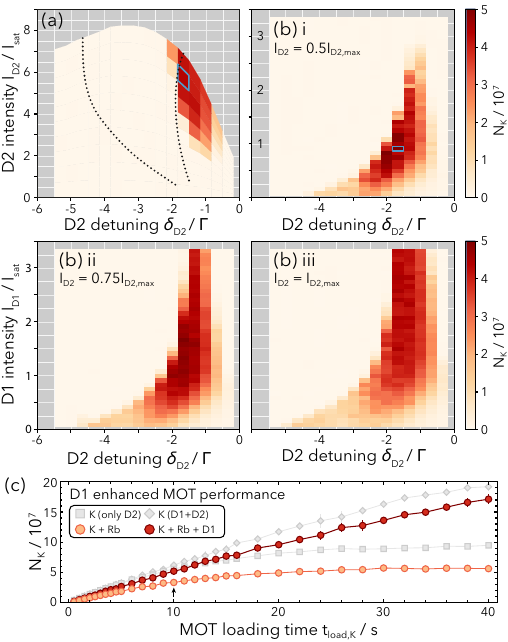}
\caption{In (a), $N_{K}$ is shown as a function of the intensity $I_{D2}$ and the frequency detuning $\delta_{D2}$ for a fixed value of $I_{D1}=\SI{1.5}{\isat}$. In (b), $N_{K}$ is plotted as a function of the intensity $I_{D1}$ and the frequency detuning $\delta_{D2}$ for a fixed value of $I_{D2}=0.5(i)/0.75(ii)/ 1.00(iii) \cdot I_{D2,max}$. In (a) and (b), the blue rectangles denote the optimal MOT operation parameters. In (c), $N_{K}$ is plotted versus the loading time $t_{load,K}$ for the respective optimal parameters of each cooling method. The optimal parameters for the two-colour cooling scheme is indicated by the blue rectangle in \autoref{KRBD1D2}(b,i). The red circles denote $N_K$ in a dual-species MOT with cooling on both the \done- and the \dtwo-transition. The orange circles denote $N_K$ in a dual-species MOT with conventional cooling only on the \dtwo-transition. The grey diamonds (squares) correspond to loading curves previously obtained in \autoref{KD1D2}(c) (\autoref{KRBD2}(c)).}
\label{KRBD1D2}
\end{figure}

\section{\label{sec:2Colour2SpeciesMOT}Two-colour, dual-species MOT}
Adding blue-detuned light on the \done transition significantly improves the dual-species MOT performance. This central result is analysed in the remainder of this article. 

In \autoref{KRBD1D2}(a), we investigate the effects of additional cooling light operating on the \done transition with an intensity of $I_{D1}=\SI{1.5}{\isat}$ in a dual-species MOT of \kk and \rb. This extends the previous investigations of \done line cooling in a single-species MOT for \kk shown in \autoref{KD1D2}. To compare our results to the previous measurements in \autoref{Konly}(b), \autoref{KD1D2}, and \autoref{KRBD2}(a), we vary the intensity $I_{D2}$ and the frequency detuning $\delta_{D2}$ and count the \kk atom number $N_{K}$. We observe a notably different behaviour throughout the entire measured parameter regime. The previously dominant \textit{conventional branch} does no longer appear. Surprisingly, we find that the inclusion of the \done cooling light restores the \textit{side branch} and even enhances the particle number $N_K$ above the saturation threshold found for the \dtwo-only dual-species MOT, as can clearly be seen in \autoref{KRBD1D2}(a). In absence of \done cooling, this parameter region is subject to significant LAHNCs and consequently low potassium particle numbers $N_{K}$. Note, that the colour map in \autoref{KRBD2} and \autoref{KRBD1D2} is equal, displaying a significantly larger \kk MOT in presence of \done cooling. In \autoref{KRBD1D2}(b), we vary the \done intensity $I_{D1}$ and the \dtwo detuning $\delta_{D2}$ for three distinct settings of the \dtwo intensity $I_{D2}$ as indicated by the insets. We observe a similar behaviour as compared to the single-species MOT and find that the performance of the MOT is sensitive to the relative powers of $I_{D1}$ and $I_{D2}$. We obtain a maximum number of $\num{5e7}$ in the case of the two-colour dual-species MOT ($I_{D1}=\SI{0.91}{\isat}$ and $\delta_{D2} = \SI{-1.67}{\Gamma}$) for a parameter setting indicated by the blue rectangle in \autoref{KRBD1D2}(b,i). This corresponds to a twofold increase for $t_{load,K} = \SI{10}{\s}$ compared to the conventional single-colour dual-species \dtwo MOT number obtained in \autoref{KRBD2}(a). Finally, we plot the potassium atom number $N_{K}$ of the dual-species MOT versus the loading time $t_{load}$ in \autoref{KRBD1D2}(c). Here, the red circles denote $N_{K}$ in the two-colour dual-species MOT. The orange circles mark $N_{K}$ for the single-colour dual-species configuration of the MOT (see \autoref{KRBD2}(c)). The grey data, for comparison, repeat the data shown in \autoref{KD1D2}(c).

For short loading times $t_{load,K}<\SI{3}{\s}$, the two cooling methods yield comparable results. However, for longer loading times $t_{load,K}>\SI{3}{\s}$ the potassium atom number $N_K$ is significantly increased by the implementation of the two-colour cooling. For the longest loading times investigated in our experiment, we find $N_K=\num{1.71e8}$ for the two-colour cooling scheme and $N_{K}=\num{5.67e7}$ for the single-colour case. This is a threefold improvement in the dual-species scenario. From the behaviour of the loading curve it is a reasonable assumption that the improvement can be even more significant for longer loading times as the \kk atom number $N_K$ in the single-colour MOT is saturated after $\sim\SI{15}{\s}$, while the reach of saturation of the potassium number in the two-colour cooling scheme is significantly shifted towards later times (cf. Appendix B).
\section{Conclusion}
In summary, we demonstrate a novel, powerful cooling scheme for preparing mixtures of cold \kk and \rb atoms in a dual-species MOT. In addition to the conventional cooling of \kk atoms using light tuned to the red side of the \dtwo-transition, a second light component, tuned to the blue side of the \done-transition, is used. We have started with benchmarking the performance of a conventional single-species MOT for \kk atoms using single frequency cooling light with a negative detuning $\delta_{D2}<0$ with respect to the \dtwo-line and an intensity $I_{D2}$. Two domains in the parameter space $\{\delta_{D2}, I_{D2} \}$ are identified, dubbed \textit{conventional branch} and \textit{side branch}, where the cooling efficiency attains a relative maximum, while the global maximum is reached in the \textit{conventional branch}. In a second step, discussed in \autoref{sec:2colourMotPotassium}, we demonstrate that the inclusion of additional light beams, blue-detuned with respect to the \done-transition, can increase the potassium atom number in a single-species MOT for long loading times. In a third step, we investigate MOT loading in a dual-species scenario of \kk and \rb. In \autoref{sec:2SpeciesMOT}, we benchmark the dual-species MOT using conventional cooling on the \dtwo-transition. Here, we find significant atomic losses due to LAHNCs and the disappearance of the \textit{side branch}. Finally, in \autoref{sec:2Colour2SpeciesMOT}, we investigate the influence of the two-colour cooling scheme in a dual-species MOT scenario. We observe a significant improvement of the \kk atom number $N_K$, with the \textit{side branch} forming the optimal parameter region. For long MOT loading times we demonstrate a threefold improvement of $N_K$. We attribute the significant reduction of saturation of \kk loading, observed in the dual-colour MOT, to a suppression of light-assisted hetero-nuclear collisions between \kk and \rb atoms. As a consequence, longer loading times or an increase of the flux of the cold \kk beam source allow for larger $N_K$. Due to the similar level structure of alkali atomic species, we suspect that the discussed cooling scheme can be a useful extension for most modern dual-species alkali quantum gas experiments, which require high fidelity preparation of large atomic samples with high repetition rates.

\section{Acknowledgements}
We acknowledge support from the Deutsche Forschungsgemeinschaft (DFG) through the collaborative research centre SFB 925 (project no. 170620586, C1). M.H. was partially supported by the Cluster of Excellence CUI: Advanced Imaging of Matter of the Deutsche Forschungsgemeinschaft (DFG)—EXC 2056—project ID 390715994.

\bibliography{apssamp}

\newpage

\appendix
\section{2D MOT parameters}\label{sec:appendix}
In the source region, separated from the main chamber by a differential pumping stage, a room temperature thermal alkali vapour is pre-cooled to produce a collimated cold atomic beam. To increase the pressure of the alkali gas, UV-LEDs are operated simultaneously with the loading phase. This method has been shown to increase the loading rates of alkali MOTs \cite{Klempt2006}. In our apparatus, the loading rate is increased almost up to tenfold using this technique \cite{Hachmann2021}. The loading rate was found to be proportional to the intensity of the UV light.

In our case, the $\textrm{2D}^+$ - MOT is formed by two pairs of counter-propagating laser beams in a retro-reflected configuration. These pairs of beams are used for transverse cooling of the particles. These four beams, together with a magnetic quadrupole field, form a conventional 2D-MOT. In our setup, however, there is an extension of this scheme, namely an additional cooling beam aligned perpendicular to the two transversal beams, called the axial cooling beam (hence the designation $\textrm{2D}^+$ - MOT). This axial beam is reflected from the surface of a substrate located inside the UHV system, which has a small hole to allow the transfer of the pre-cooled atoms into the main chamber through a differential pumping stage. The $\textrm{2D}^+$ - MOT operates only on the \dtwo-line of \kk and therefore addresses the $\ket{F=9/2}\rightarrow\ket{F'=11/2}$ transition for cooling and the$\ket{F=7/2}\rightarrow\ket{F'=9/2}$ transition for repumping of the atoms (see \autoref{Konly}(a) in main text). The magnetic quadrupole field to generate the transversal magnetic confinement is provided by two pairs of coils placed outside the UHV source region. Static stray magnetic fields are compensated by additional coil pairs in the transversal direction. This magnetic field infrastructure allows for magnetic gradients up to 20 G/cm and homogeneous magnetic compensation fields on the order of several \si{\gauss}. The experimentally optimised parameters of the $\textrm{2D}^+$ - MOT can be found in \autoref{tab:table2DK}. Efficient operation of the $\textrm{2D}^+$ - MOT is a crucial step in the experiments, since the atomic flux determined by the quality of the $\textrm{2D}^+$ - MOT strongly influences the performance of the 3D-MOT and thus all subsequent steps in the production of cold atoms.
\begin{table}[b]
\begin{ruledtabular}
\begin{tabular}{cll}
\multicolumn{1}{c}{\textrm{Parameter}}&\textrm{Transversal}&\textrm{Axial}\\
\colrule
magnetic gradient & \SI{17.5}{\gauss\per\centi\meter} & \SI{0}{\gauss\per\centi\meter} \\
1/e beam diameter & \SI{26.9}{\milli\meter} & \SI{14.4}{\milli\meter} \\
$I_{cool}$ per beam & \SI{28}{\isat} & \SI{70}{\isat} \\
$I_{rep}$ per beam & \SI{27.5}{\isat} & \SI{14}{\isat} \\
$\delta_{cool}$ & \SI{-3}{\Gamma} & \SI{-3.6}{\Gamma} \\
$\delta_{rep}$ & \SI{-0.05}{\Gamma} & \SI{-0.05}{\Gamma} \\
\end{tabular}
\end{ruledtabular}
\caption{\label{tab:table2DK}
Experimentally optimised \kk 2D+-MOT parameters.
}
\end{table}
\begin{table}
\begin{ruledtabular}
\begin{tabular}{cll}
\multicolumn{1}{c}{\textrm{Parameter}}&\textrm{Transversal}&\textrm{Axial}\\
\colrule
magnetic gradient & \SI{14}{\gauss\per\centi\meter} & \SI{0}{\gauss\per\centi\meter} \\
1/e beam diameter & \SI{26.9}{\milli\meter} & \SI{10.8}{\milli\meter} \\
$I_{cool}$ per beam& \SI{8}{\isat} & \SI{20}{\isat} \\
$I_{rep}$ per beam& \SI{0.3}{\isat} & \SI{0}{\isat} \\
$\delta_{cool}$ & \SI{-2.7}{\Gamma} & \SI{-2.7}{\Gamma} \\
$\delta_{rep}$ & \SI{0}{\Gamma} & -\\
\end{tabular}
\end{ruledtabular}
\caption{\label{tab:table2DRb}
Experimentally optimised \rb 2D+-MOT parameters.
}
\end{table}

The rubidium $\textrm{2D}^+$ - MOT is realised in the same configuration as the potassium $\textrm{2D}^+$ - MOT. The experimentally optimised parameters of the \rb $\textrm{2D}^+$ - MOT can be found in \autoref{tab:table2DRb}. The experimentally optimised parameters of the \rb and \kk 3D-MOT can be found in \autoref{tab:table3DMOT}. The parameter values presented in the appendix of this work are not varied throughout the series of experiments described in the main text.

\begin{table}
\begin{ruledtabular}
\begin{tabular}{cl}
\multicolumn{1}{c}{\textrm{Parameter}}&\textrm{Transversal}\\
\colrule
magnetic gradient & \SI{7.5}{\gauss\per\centi\meter}\\
1/e beam diameter & \SI{26.9}{\milli\meter} \\
$I_{cool}$ per beam & \SI{1.7}{\isat}\\
$I_{rep}$ per beam& \SI{0.04}{\isat}\\
$\delta_{cool}$ & \SI{-2.7}{\Gamma}\\
$\delta_{rep}$ & \SI{0}{\Gamma}\\
\end{tabular}
\end{ruledtabular}
\caption{\label{tab:table3DMOT}
Experimentally optimised \rb 3D-MOT parameters.
}
\end{table}

\section{Long-term MOT loading behaviour}\label{sec:appendixLoadingCurves}

\begin{figure*}
\includegraphics[width=2\columnwidth]{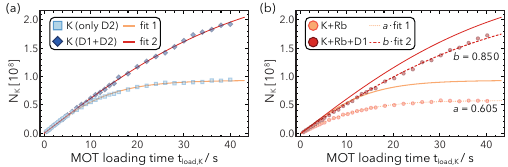}
\caption{In (a), we repeat the data obtained in the main text of this work of the single-species MOT (\autoref{KD1D2}) and plot the potassium atom number $N_K$ as a function of the loading time $t_{load,K}$. In (a), the salmon and red solid line denotes the fit of the single-species data according to the correct model for the single-colour and two-colour MOT, respectively. For details, see text in Appendix \ref{sec:appendixLoadingCurves}. In (b), we repeat the data obtained in the main text of the dual-species MOT (\autoref{KRBD1D2}) and plot the potassium atom number $N_K$ as a function of the loading time $t_{load,K}$. Here, the dotted lines denote  a fit according to the insets, where the curves fit1 and fit2 are scaled by fitted factors $a = 0.605$ (salmon) and $b=0.850$ (red), respectively. }
\label{fig:loadingCurvesFitting}
\end{figure*} 

To gain insights about the long-term loading behaviour of the different MOT loading scenarios, we fit our experimental data obtained in the main text (\autoref{KD1D2}(c) and \autoref{KRBD1D2}(c)) to a rate equation model commonly utilised describing the loading behaviour of single-species MOTs\cite{Dinneen1999,Busch2006}: 
\begin{equation}
    \frac{\partial N_x(t)}{\partial t} = R - \gamma N_x(t) - \beta_x N_x(t)^2
\end{equation}
where $N_x(t)$ denotes the particle number $N_K$ found in the single-colour (x = D2) of dual-colour (x = D1D2) operation. We extract the constant atomic loading rate R and the single-particle loss rate $\gamma$ with a simultaneous fit to both loading curves shown in \autoref{fig:loadingCurvesFitting} (a). The additional free parameter $\beta_x$ models the density-dependent loss rate, which is assumed to be different between the two scenarios and is therefore individually determined. We find, that the ratio of the density-dependent loss rate $\beta_{D2}/\beta_{D1D2} = 9.73$ is far from unity, resulting in a vastly different saturation level in both cases. Extrapolating these loading curves beyond the measured loading duration leads to a difference in the saturation level of \num{2.96}. 

In the dual-species setting, isolating individual processes is much more challenging and therefore the loading behaviour can no longer be captured with a simple rate equation model. The inclusion of an additional inter-species density-dependent loss term fails to accurately describe the experimental data. In \autoref{fig:loadingCurvesFitting} (b), we plot the potassium atom number $N_K$ of the dual-species MOT as a function of the loading time $t_{load,K}$. The solid lines repeat the fits obtained in (a) for comparability. Interestingly we find, that the data is best modeled by rescaling the previously determined fit functions fit 1, fit 2 with factors $a = \num{0.605}$ and $b = \num{0.850}$, respectively. Once again, extrapolating beyond the measured data, we obtain a difference in saturation level of $\num{2.96} * a/b = \num{4.15}$. However, without access to the correct rate equation models, this result has to be taken with care.

\end{document}